\documentclass[preprint]{ptephy_v1}

\preprintnumber{IU-TH-19} 
\usepackage{hyperref}

\usepackage{amsmath}
\usepackage{amssymb}

\usepackage{graphicx}
\usepackage{color}

\usepackage{float}



\newcommand{\tr}{\mathop{\rm tr}\nolimits}

\begin{document}

\title{
Automatic hermiticity for mixed states
}


\author{Keiichi Nagao${}^{1,2,*}$} 

\author{Holger Bech Nielsen${}^{2,*}$}

\affil{${}^{1}$Faculty of Education, Ibaraki University, Bunkyo 2-1-1, Mito 
310-8512, Japan }

\affil{${}^{2}$Niels Bohr Institute, University of Copenhagen, 
Blegdamsvej 17, Copenhagen $\O$ 2100, Denmark}

\affil{
\email{keiichi.nagao.phys@vc.ibaraki.ac.jp; hbech@nbi.dk}
}

%


\begin{abstract}%

We previously proposed a mechanism to effectively obtain, after a long time development, 
a Hamiltonian being 
Hermitian with regard to a modified inner product $I_Q$ that makes 
a given non-normal Hamiltonian 
normal by using an appropriately chosen Hermitian operator $Q$. 
We studied it for pure states. 
In this letter we show that a similar mechanism also works for mixed states 
by introducing density matrices to describe them and 
investigating their properties explicitly both in the future-not-included and future-included theories. 
In particular, in the latter, 
where not only a past state at the initial time $T_A$ but also a future state at the final time $T_B$ 
is given, 
we study a couple of candidates for it, 
and introduce a ``skew density matrix'' composed of both ensembles of the future and past states  
such that the trace of the product of it and an operator ${\cal O}$ matches 
a normalized matrix element of ${\cal O}$.   
We argue that 
the skew density matrix defined with $I_Q$ at the present time $t$ 
for large $T_B-t$ and large $t-T_A$ approximately corresponds to 
another density matrix composed of only an ensemble of past states 
and defined with 
another inner product 
$I_{Q_J}$ for large $t-T_A$.

\end{abstract}


\maketitle

%

\section{Introduction}

Quantum theory is described well via the Feynman path integral. 
In the Feynman path integral, the action in the integrand is considered to be path-dependent, 
while the measure is usually supposed to be path-independent. 
However, we could consider a theory such that not only the action but also the measure is path-dependent. 
This is the complex action theory (CAT) 
whose action is complex at a fundamental level but expected to look real effectively\cite{Bled2006}. 
The CAT provides us with 
falsifiable predictions\cite{Bled2006,Nielsen:2008cm,Nielsen:2007ak,Nielsen:2005ub}. 
Deeper understanding via the CAT have been tried even for the Higgs mass\cite{Nielsen:2007mj}, 
quantum mechanical philosophy\cite{newer1,Vaxjo2009,newer2}, 
some fine-tuning problems\cite{Nielsen2010qq,degenerate}, 
black holes\cite{Nielsen2009hq}, 
de Broglie-Bohm particles and a cut-off in loop diagrams\cite{Bled2010B}, 
a mechanism to obtain Hermitian Hamiltonians\cite{Nagao:2010xu}, 
the complex coordinate formalism\cite{Nagao:2011za}, 
and the momentum relation\cite{Nagao:2011is,Nagao:2013eda}. 
There are two types of CAT. 
One is the future-not-included theory, where 
only the past state $| A(T_A) \rangle$ at the initial time $T_A$ is given 
and the time integration is performed over the period from $T_A$ to 
a reference time $t$. 
The other is the future-included one, where not only the past state $| A(T_A) \rangle$ 
but also the future state $| B(T_B) \rangle$ at the final time $T_B$ is given, 
and the time integration is performed over the whole period from $T_A$ to $T_B$. 
We elucidated various interesting properties of the future-not-included CAT\cite{Nagao:2013eda}.  
In Ref.\cite{Nagao:2017ecx} we argued that, if a theory is described 
with a complex action, then such a theory is suggested to be the future-included theory, 
rather than the future-not-included one, as long as we respect objectivity. 
Even so, the future-not-included CAT itself still remains a fascinating theory, 
and a good playground to study various intriguing aspects of the CAT.

In the future-included theory, 
the normalized matrix element  
$\langle \hat{\mathcal O} \rangle^{BA} 
\equiv \frac{ \langle B(t) |  \hat{\mathcal O}  | A(t) \rangle }{ \langle B(t) | A(t) \rangle }$ 
of an operator $\hat{\mathcal O}$ is expected to have the role of 
an expectation value\cite{Bled2006}.\footnote{The normalized matrix element $\langle \hat{\mathcal O} \rangle^{BA}$ is called 
the weak value\cite{AAV,generalized_two-state_vector_formalism} and has been studied intensively with a different philosophy 
in the real action theory (RAT). For details, see Ref.\cite{review_wv} and references therein.} 
Indeed, if we regard it so, 
we can obtain nice properties such as the Heisenberg equation, Ehrenfest's theorem, 
and a conserved probability current density\cite{Nagao:2012mj,Nagao:2012ye}. 
However, $\langle \hat{\mathcal O} \rangle^{BA}$ is generically complex even 
for Hermitian $\hat{\mathcal O}$, even though 
any observables are real. 
To resolve this problem, in Refs.\cite{Nagao:2015bya,Nagao:2017cpl}, 
we proposed a theorem that states that, 
provided that an operator $\hat{\mathcal O}$ is $Q$-Hermitian, i.e., 
Hermitian with regard to a modified inner product $I_Q$ that makes a given non-normal 
Hamiltonian\footnote{The Hamiltonian $\hat{H}$ is 
generically non-normal, so it is 
not restricted to the class of PT-symmetric non-Hermitian Hamiltonians that were 
studied in 
Refs.\cite{Bender:1998ke,Bender:1998gh,Mostafazadeh_CPT_ip_2002,Mostafazadeh_CPT_ip_2003,Bender:2011ke}.} normal 
by using an appropriately chosen Hermitian operator $Q$, 
the normalized matrix element defined with $I_Q$ becomes real and 
time-develops under a $Q$-Hermitian Hamiltonian for the past and future states selected 
such that the absolute value of the transition amplitude defined with $I_Q$ from the past state 
to the future state is maximized. 
We call this way of thinking the maximization principle. 
We proved the theorem in the case of non-normal 
Hamiltonians $\hat{H}$\cite{Nagao:2015bya}\footnote{The proof is based on the existence of imaginary parts of the eigenvalues of $\hat{H}$, so it cannot be applicable to the RAT case. 
The maximization principle is reviewed in Refs.\cite{Nagao:2017book,Nagao:2017ztx}.} and 
in the real action theory (RAT)\cite{Nagao:2017cpl}. 
In addition, we studied the periodic CAT and 
proposed a variant type of the maximization principle, 
by which the period could be determined\cite{Nagao:2022rap}.

The maximization principle is based on the natural way of thinking and looks promising. 
Behind the principle, the automatic hermiticity mechanism\cite{Nagao:2010xu} has a key role. 
In the CAT the imaginary parts of the eigenvalues $\lambda_i$ 
of a given non-normal Hamiltonian $\hat{H}$ are supposed to be bounded from above for 
the Feynman path integral $\int e^{\frac{i}{\hbar}S} {\mathcal D} \text{path}$ to converge. 
Then we can imagine that some $\text{Im} \lambda_i$ take the maximal value $B$, and
denote the corresponding subset of $\{ i \}$ as $A$. 
After a long time development, only the subset $A$ contributes most significantly, 
and on the subset a $Q$-Hermitian Hamiltonian effectively emerges . 
This is the automatic hermiticity mechanism that we proposed and studied explicitly 
for pure states time-developing forward\cite{Nagao:2010xu}. 
In Ref.\cite{Nagao:2012mj}, utilizing it for pure states time-developing forward and backward, 
we showed that the normalized matrix element of ${\cal O}$ 
at the present time $t$ 
in the future-included theory for large $T_B-t$ and large $t- T_A$ corresponds to 
the expectation value of ${\cal O}$ in the future-not-included theory 
defined with a modified inner product $I_{Q'}$ for large $t- T_A$. 
This study strongly suggests that the future-included CAT is not excluded phenomenologically, 
even though it looks very exotic. 
The automatic hermiticity mechanism has an essential role for the CAT to be viable, 
but so far we have studied it only for pure states, not for mixed states.  
Thus it would be natural to pose the question: how does it work for mixed states? 
Even though mixed states can always be expressed by pure states 
defined in a larger system that includes the mixed states in its subsystem, 
it is interesting and worthwhile to study how mixed states are defined and how they 
behave in the CAT. 
In particular, it is intriguing to study the automatic hermiticity mechanism for mixed states 
in the CAT, because the emergence of a Hermitian Hamiltonian is crucially important for the CAT 
to be sensible, and also because mixed states are generic quantum states along with pure states.

We need to introduce density matrices to describe mixed states in the CAT. 
In the future-not-included CAT, there is only 
one class of state vectors 
time-developing forward from the past, 
while in the future-included CAT there are two classes of state vectors 
time-developing not only forward 
but also backward from the future. 
Hence it would be more non-trivial to define density matrices 
and see the emergence of hermiticity for them in the future-included CAT 
rather than the future-not-included one. 
Therefore, in this letter, after reviewing the modified inner product $I_Q$ and automatic hermiticity mechanism for pure states, 
we first define density matrices to describe mixed states 
and study the emergence of hermiticity for them in the future-not-included CAT. 
Next, we investigate 
a couple of candidates for density matrices in the future-included CAT, 
and introduce a ``skew density matrix'' composed of both ensembles of the future and past states 
such that the trace of the product of it and an operator ${\cal O}$ becomes  
a normalized matrix element of ${\cal O}$. 
Furthermore, we argue that the skew density matrix defined with $I_Q$ at the present time $t$ 
for large $T_B-t$ and large $t-T_A$ approximately corresponds to 
another type of density matrix composed of only an ensemble of the past state
and defined with another inner product $I_{Q_J}$ for large $t-T_A$. 
Finally we summarize the study in this letter and discuss the outlook of our theory.

\section{Modified inner product and the automatic hermiticity mechanism for pure states}

In this section we briefly review the modified inner product $I_Q$ and 
the automatic hermiticity mechanism 
for pure states by following Refs.\cite{Nagao:2010xu, Nagao:2011za}. 
The eigenstates of a given non-normal Hamiltonian $\hat{H}$, 
$| \lambda_i \rangle (i=1,2,\dots)$ 
obeying $\hat{H} | \lambda_i \rangle = \lambda_i | \lambda_i \rangle$,  are not orthogonal to 
each other in the usual inner product $I$. 
Let us introduce a modified inner product 
$I_Q$\cite{Nagao:2010xu,Nagao:2011za}\footnote{Similar inner products are also studied 
in Refs.\cite{Geyer,Mostafazadeh_CPT_ip_2002,Mostafazadeh_CPT_ip_2003}.} such that  
$| \lambda_i \rangle (i=1,2,\dots)$ become orthogonal to each other 
with regard to it, i.e., 
for arbitrary kets $|u \rangle$ and $|v \rangle$, $I_Q(|u \rangle , |v \rangle) \equiv \langle u |_Q v \rangle 
\equiv \langle u | Q | v \rangle$, 
where $Q$ is a Hermitian operator that obeys $\langle \lambda_i  |_Q \lambda_j \rangle = \delta_{ij}$. 
Using the diagonalizing operator $P=(| \lambda_1 \rangle , | \lambda_2 \rangle , \ldots)$ 
such that $P^{-1}\hat{H}P=D=\text{diag}(\lambda_1, \lambda_2, \dots)$, 
we choose $Q=(P^\dag)^{-1} P^{-1}$. 
Next we define the $Q$-Hermitian conjugate $\dag^Q$ of 
an operator $A$ by $\langle \psi_2 |_Q A | \psi_1 \rangle^* \equiv \langle \psi_1 |_Q A^{\dag^Q} | \psi_2 \rangle$, so $A^{\dag^Q} \equiv Q^{-1} A^\dag Q$. 
If $A^{\dag^Q} = A$, we call $A$ $Q$-Hermitian. 
We also introduce $\dag^Q$ for kets and bras as 
$| \lambda \rangle^{\dag^Q} \equiv \langle \lambda |_Q $ and 
$\left(\langle \lambda |_Q \right)^{\dag^Q} \equiv | \lambda \rangle$. 
Then, since $P^{-1}=P^{\dag^Q}$, $\hat{H}$ is $Q$-normal, $[\hat{H}, \hat{H}^{\dag^Q} ] =0$. 
We can decompose $\hat{H}$ as 
$\hat{H}=\hat{H}_{Qh} + \hat{H}_{Qa}$, 
where $\hat{H}_{Qh}= \frac{\hat{H} + \hat{H}^{\dag^Q} }{2}$ and 
$\hat{H}_{Qa} = \frac{\hat{H} - \hat{H}^{\dag^Q} }{2}$ are 
$Q$-Hermitian and anti-$Q$-Hermitian parts of $\hat{H}$, respectively.


Let us consider a state\footnote{We consider a state with an index $i$ just for our later convenience. } 
$|A_i (t) \rangle$, which obeys the Schr\"{o}dinger equation 
\begin{equation}
i \hbar \frac{d}{dt} | A_i (t) \rangle = \hat{H} | A_i (t) \rangle. \label{Schr_for_Ai} 
\end{equation} 
We introduce a normalized state and an expectation value of an operator ${\cal O}$ by 
$| A_i (t) \rangle_{N} 
\equiv \frac{1}{\sqrt{ \langle A_i (t) |_Q ~{A_i}(t) \rangle} } | {A_i}(t) \rangle$, 
$\langle \hat{\cal O} \rangle_Q^{A_i A_i} (t) 
\equiv  {}_{N} \langle A_i (t) |_Q {\cal O} | A_i (t) \rangle_{N}$. 
They obey 
$i\hbar \frac{\partial}{ \partial t} | A_i (t) \rangle_{N} 
= \hat{H}_{Qh} | A_i (t) \rangle_{N} 
+ \hat{\Delta} \left( \hat{H}_{Qa} ; | A_i (t) \rangle_{N} \right) 
| A_i (t) \rangle_{N}$, 
$\frac{d}{dt}\langle \hat{\cal O} \rangle_Q^{A_i A_i} (t)
= 
-\frac{i}{\hbar} \langle \left[ \hat{\cal O} , \hat{H}_{Qh} \right] \rangle_Q^{A_i A_i} (t) 
-\frac{i}{\hbar} \langle  \left\{  \hat{\cal O}, \hat{\Delta} \left( \hat{H}_{Qa} ; | A_i (t) \rangle_{N} \right) \right\} \rangle_Q^{A_i A_i} (t)$, 
where we have introduced 
$\hat{\Delta} \left( \hat{H}_{Qa} ; | A_i (t) \rangle_{N} \right) 
\equiv \hat{H}_{Qa} -{}_{N} \langle A_i (t) |_Q \hat{H}_{Qa} | A_i (t) \rangle_{N}$. 
It seems that, in the classical limit, 
since 
$\langle  \left\{  \hat{\cal O}, \hat{\Delta} \left( \hat{H}_{Qa} ; | A_i (t) \rangle_{N} \right) \right\} \rangle_Q^{A_i A_i} (t)$ 
is suppressed, 
$\langle \hat{\cal O} \rangle_Q^{A_i A_i}(t)$ 
time-develops by a $Q$-Hermitian Hamiltonian, and Ehrenfest's theorem holds. 
This property is intriguing, but we will see the emergence of the $Q$-hermiticity 
even before considering the classical limit via the automatic hermiticity mechanism, which we explain below.

Expanding $| A_i (t) \rangle$ as 
$| A_i (t) \rangle = \sum_j a_j^{(i)} (t) | \lambda_j \rangle$ 
and introducing $| A'_i (t) \rangle = P^{-1} | A_i (t) \rangle= \sum_j a_j^{(i)} (t) | e_j \rangle$, 
which obeys $i \hbar \frac{d}{dt} | A'_i(t) \rangle =D | A'_i(t) \rangle$, 
we obtain 
$| A_i(t) \rangle = P e^{- \frac{i}{\hbar} D (t-t_0)} | A'_i(t_0) \rangle 
= \sum_j a_j^{(i)} (t_0) 
e^{ \frac{1}{\hbar} \left( \text{Im} \lambda_j - i \text{Re} \lambda_j \right) (t-t_0)}       
| \lambda_j \rangle$. 
Now we assume that the anti-$Q$-Hermitian part of $\hat{H}$ is bounded from above 
for the Feynman path integral $\int e^{\frac{i}{\hbar}S} {\cal D}\text{path}$ to converge. 
Based on this assumption 
we can imagine that some $\text{Im} \lambda_j$ 
take the maximal value $B$, and denote the corresponding subset of $\{ j \}$ as $A$. 
After a long time has passed, i.e., for large $t-t_0$, 
the states with $\text{Im} \lambda_j |_{j \in A}$ 
contribute most in the sum.  
Let us define a diagonalized Hamiltonian $\tilde{D}_{R}$ by 
\begin{equation}
\langle e_j | \tilde{D}_{R} | e_k \rangle \equiv 
\left\{ 
 \begin{array}{cc}
      \langle e_j | D_R | e_k \rangle =\delta_{jk} \text{Re} \lambda_j  & \text{for} \quad j \in A , \\
      0 &\text{for} \quad j \not\in A , \\ 
 \end{array}
\right. \label{DRtilder}
\end{equation}
and introduce $\hat{H}_{\text{eff}} \equiv P \tilde{D}_{R} P^{-1}$. 
Since $(\tilde{D}_{R})^{\dag} = \tilde{D}_{R}$, $\hat{H}_{\text{eff}}$ is $Q$-Hermitian, 
$\hat{H}_{\text{eff}} ^{\dag^Q} =\hat{H}_{\text{eff}}$, and  
satisfies $\hat{H}_{\text{eff}} | \lambda_i \rangle = \text{Re} \lambda_i | \lambda_i \rangle$. 
Then $| A_i (t) \rangle$ is evaluated as 
$| A_i (t) \rangle 
\simeq e^{ \frac{1}{\hbar} B (t-t_0)} 
\sum_{j \in A}  a_j^{(i)} (t_0)$ 
$e^{-\frac{i}{\hbar} {\text Re} \lambda_j (t-t_0)} | \lambda_j \rangle 
=e^{ \frac{1}{\hbar} B (t-t_0)}  e^{-\frac{i}{\hbar} \hat{H}_{\text{eff}} (t-t_0)} 
| \tilde{A}_i (t_0) \rangle = | \tilde{A}_i (t) \rangle$, 
where we have introduced $| \tilde{A}_i (t) \rangle \equiv \sum_{j \in A}  a_j^{(i)} (t)| \lambda_j \rangle$. 
Since the factor $e^{ \frac{1}{\hbar} B (t-t_0)}$ 
is dropped out for the normalized state $| A_i(t) \rangle_N$, 
we have effectively obtained a $Q$-Hermitian Hamiltonian $\hat{H}_{\text{eff}}$ 
after the long time development. 
In fact, the normalized state 
$| A_i (t) \rangle_{N} 
\simeq 
| \tilde{A}_i (t) \rangle_{N}
\equiv
\frac{1}{\sqrt{ \langle \tilde{A}_i (t) |_Q ~\tilde{A}_i(t) \rangle} } | \tilde{A}_i (t) \rangle$ 
and the expectation value of an operator ${\cal O}$, 
$\langle \hat{\cal O} \rangle_Q^{A_i A_i} (t) 
\simeq 
\langle \hat{\cal O} \rangle_Q^{\tilde{A}_i \tilde{A}_i} (t) 
\equiv 
{}_{N} \langle \tilde{A}_i (t) |_Q {\cal O} | \tilde{A}_i (t) 
\rangle_{N}$, 
obey 
\begin{eqnarray}
&&i\hbar \frac{\partial}{ \partial t} | \tilde{A}_i (t) \rangle_{N} = \hat{H}_{\text{eff}} | \tilde{A}_i (t) \rangle_{N}, \label{Schreq_for_AitildeketN} \\
&&
\frac{d}{dt} \langle \hat{\cal O} \rangle_Q^{\tilde{A}_i \tilde{A}_i} (t) 
=
-\frac{i}{\hbar}  \langle  \left[  \hat{\cal O}, \hat{H}_{\text{eff}} \right] 
\rangle_Q^{\tilde{A}_i \tilde{A}_i} (t).  
\end{eqnarray}
Thus we have seen for pure states that 
the $Q$-hermitian Hamiltonian $\hat{H}_{\text{eff}}$ emerges.


\section{Density matrices for mixed states in the future-not-included CAT}

In this section we define density matrices to describe mixed states 
and study the automatic hermiticity mechanism for them in the future-not-included CAT. 
For a given ensemble\footnote{We note that each $| A_i (t) \rangle$ does not need to be orthogonal to each other, and that the number of elements does not have to match the order of the Hilbert space. } 
$\left\{ | A_i (t) \rangle \right\}$, 
each of which obeys Eq.(\ref{Schr_for_Ai}), 
let us consider a mixed state that is composed of $| A_i (t) \rangle_{N}$ 
with the probability $q_i$ for each index $i$ ($q_i \ge 0$, $\sum_i q_i=1$).  
We define the density matrix\footnote{When the density matrix is composed of only one component, 
$\hat{\rho}_Q^{A,\text{mixed}}(t)=| A_1(t) \rangle_N {}_N \langle A_1(t) |_Q$, 
it describes a pure state, 
and satisfies $\hat{\rho}_Q^{A,\text{mixed}}(t)^2 = \hat{\rho}_Q^{A,\text{mixed}}(t)$ and 
$\tr\left(\hat{\rho}_Q^{A,\text{mixed}}(t)\right)=1$.} 
and expectation value of an operator $\hat{\cal O}$ for it by 
\begin{eqnarray}
\hat{\rho}_Q^{AA,\text{mixed}}(t) 
&\equiv&  \sum_i q_i  | A_i (t) \rangle_N {}_N \langle A_i (t) |_Q 
\equiv \sum_i q_i \hat{\rho}_Q^{A_i A_i}(t) ,\label{rhoQAmixed} \\
\langle \hat{\cal O} \rangle_{\hat{\rho}_Q^{AA,\text{mixed}}} (t)
&\equiv&
\tr\left( \hat{\rho}_Q^{AA,\text{mixed}}(t)  \hat{\cal O}   \right) 
\equiv \sum_i q_i \langle \hat{\cal O} \rangle_{\hat{\rho}_Q^{A_i A_i}} (t)  
=\sum_i q_i \langle \hat{\cal O} \rangle_Q^{A_i A_i} (t) , 
\end{eqnarray}
where $\hat{\rho}_Q^{A_i A_i} (t)$ obeys  
$\hat{\rho}_Q^{A_i A_i}(t)^2 = \hat{\rho}_Q^{A_i A_i}(t)$ 
and $\tr\left(\hat{\rho}_Q^{A_i A_i} (t) \right)=1$. 
So $\tr\left(\hat{\rho}_Q^{A A,\text{mixed}}(t) \right)=1$. 
We note that $\hat{\rho}_Q^{A A,\text{mixed}}(t)$ and 
$\langle \hat{\cal O} \rangle_{\hat{\rho}_Q^{AA,\text{mixed}}} (t)$ 
are $Q$-Hermitian and real for $Q$-Hermitian $\hat{\cal O}$, respectively. 
They time-develop as follows: 
$\frac{d}{dt} \hat{\rho}_Q^{AA,\text{mixed}} (t) 
=-\frac{i}{\hbar} \left[ \hat{H}_{Qh}, \hat{\rho}_Q^{AA,\text{mixed}} (t) \right] 
-\frac{i}{\hbar} \sum_i q_i 
\left\{ \hat{\Delta} \left( \hat{H}_{Qa} ;  | A_i (t) \rangle_{N} \right), 
\hat{\rho}_Q^{A_i A_i}  \right\} (t)$, 
$\frac{d}{dt} \langle \hat{\cal O} \rangle_{\hat{\rho}_Q^{AA,\text{mixed}}} (t) 
= 
-\frac{i}{\hbar} \langle \left[ \hat{\cal O} , \hat{H}_{Qh} \right] \rangle_{\hat{\rho}_Q^{AA,\text{mixed}}} (t)$ 
$-\frac{i}{\hbar} \sum_i q_i$ 
$\langle  \left\{  \hat{\cal O}, \hat{\Delta} \left( \hat{H}_{Qa} ; | A_i (t) \rangle_{N} \right)  \right\} \rangle_{\hat{\rho}_Q^{A_i A_i}} (t)$. 
It is interesting that, in the classical limit, 
since $\langle  \left\{  \hat{\cal O}, \hat{\Delta} \left( \hat{H}_{Qa} ; | A_i (t) \rangle_{N} \right)  \right\} \rangle_{\hat{\rho}_Q^{A_i A_i}} (t)$ is suppressed, 
Ehrenfest's theorem holds. 
Now, let us consider the long time development. 
Then, since $| A_i (t) \rangle_{N} \simeq | \tilde{A}_i (t) \rangle_{N}$ 
obeys Eq.(\ref{Schreq_for_AitildeketN}), 
we obtain the following relations 
for $\hat{\rho}_Q^{AA,\text{mixed}} (t) \simeq \hat{\rho}_Q^{\tilde{A} \tilde{A}, \text{mixed}} (t)$, 
$\hat{\rho}_Q^{A_i A_i} (t) \simeq \hat{\rho}_Q^{\tilde{A}_i \tilde{A}_i}(t)$, 
$\langle \hat{\cal O} \rangle_{\hat{\rho}_Q^{AA,\text{mixed}}} (t)
\simeq \langle \hat{\cal O} \rangle_{\hat{\rho}_Q^{\tilde{A} \tilde{A},\text{mixed}}} (t)$, 
and $\langle \hat{\cal O} \rangle_{\hat{\rho}_Q^{A_i A_i}} (t) \simeq \langle \hat{\cal O} \rangle_{\hat{\rho}_Q^{\tilde{A}_i \tilde{A}_i}}(t)$: 
\begin{eqnarray}
&&
\hat{\rho}_Q^{\tilde{A} \tilde{A}, \text{mixed}} (t) 
\equiv \sum_{i} q_i \hat{\rho}_Q^{\tilde{A}_i \tilde{A}_i}(t) 
=
\hat{U}_{\text{eff}} (t-T_A) 
\hat{\rho}_Q^{\tilde{A} \tilde{A},\text{mixed}}(T_A) 
\hat{U}_{\text{eff}} (t-T_A)^{\dag^Q} , \label{rhohatAtildemixedQt} \\
&&
\langle \hat{\cal O} \rangle_{\hat{\rho}_Q^{\tilde{A} \tilde{A},\text{mixed}}} (t) 
\equiv
\tr\left( \hat{\rho}_Q^{\tilde{A} \tilde{A},\text{mixed}}(t)  \hat{\cal O}   \right) 
\equiv \sum_i q_i \langle \hat{\cal O} \rangle_{\hat{\rho}_Q^{\tilde{A}_i \tilde{A}_i}} (t) 
=\sum_i q_i \langle \hat{\cal O} \rangle_Q^{\tilde{A}_i \tilde{A}_i} (t) , \\
&&
\frac{d}{dt} \hat{\rho}_Q^{\tilde{A} \tilde{A},\text{mixed}}(t)   
= 
-\frac{i}{\hbar} \left[ \hat{H}_{\text{eff}} , \hat{\rho}_Q^{\tilde{A} \tilde{A},\text{mixed}}(t)  \right] , \\
&&
\frac{d}{dt} \langle \hat{\cal O} \rangle_{\hat{\rho}_Q^{\tilde{A} \tilde{A},\text{mixed}}} (t)
=
-\frac{i}{\hbar}  \langle  \left[  \hat{\cal O}, \hat{H}_{\text{eff}} \right] 
\rangle_{\hat{\rho}_Q^{\tilde{A} \tilde{A},\text{mixed}}} (t) ,  
\end{eqnarray}
where $\hat{U}_{\text{eff}} (t-T_A) \equiv e^{-\frac{i}{\hbar} \hat{H}_{\text{eff}} (t-T_A)}$ 
is ``$Q$-unitary'', 
i.e., $U_{\text{eff}} (t-T_A)^{\dag^Q}=U_{\text{eff}} (t-T_A)^{-1}$.
We find that $\hat{\rho}_Q^{\tilde{A}\tilde{A}, \text{mixed}} (t)$ 
obeys the von Neumann equation with 
the $Q$-Hermitian Hamiltonian $\hat{H}_{\text{eff}}$ and Ehrenfest's theorem holds. 
Thus we have confirmed that the automatic hermiticity mechanism works for mixed states as well as 
for pure states in the future-not-included CAT.


\section{Density matrices for mixed states in the future-included CAT}

In this section we attempt to introduce density matrices to describe mixed states 
and study their properties in the future-included CAT. 
In addition we investigate the automatic hermiticity mechanism for the mixed states. 
The future-included theory is described not only by 
the state vector $| A_i (t) \rangle$ that time-develops forward from the initial time $T_A$ 
according to 
the Schr\"{o}dinger equation (\ref{Schr_for_Ai}) but also by 
the other one\footnote{We adopt a convention with an index $i$ for the state, 
since it will be convenient later.} 
$| B_i (t) \rangle$ that time-develops backward from the final time $T_B$ according to 
the other Schr\"{o}dinger equation:\footnote{In Ref.\cite{Nagao:2015bya}, we adopted the modified inner product $I_Q$ 
for all quantities in the future-included CAT\cite{Bled2006,Nagao:2012mj,Nagao:2012ye}. We follow this formalism in this letter.}  
$i \hbar \frac{d}{dt} | B_i (t) \rangle = {\hat{H}}^{\dag^Q} | B _i (t) \rangle 
\Leftrightarrow -i \hbar \frac{d}{dt} \langle B_i (t) |_Q = \langle B_i (t) |_Q  \hat{H}$. 
The states $| A_i (t) \rangle$ and $| B_i (t) \rangle$ are normalized by 
$\langle A_i (T_A) |_Q A_i (T_A) \rangle = \langle B_i (T_B) |_Q B_i (T_B) \rangle = 1$. 
The normalized matrix element\footnote{In the case of $Q=1$, this corresponds to the weak value\cite{AAV, review_wv} that is well known in the RAT.} 
\begin{equation}
\langle \hat{\mathcal O} \rangle_Q^{B_i A_i} (t) 
\equiv \frac{ \langle B_i(t) |_Q  \hat{\mathcal O}  | A_i(t) \rangle }{ \langle B_i(t) |_Q A_i(t) \rangle } 
\label{OQB_iA_i}
\end{equation} 
is a good candidate for an expectation value of an operator ${\cal O}$ in the future-included CAT, 
because, if it is viewed as such, 
then we can obtain the Heisenberg equation, Ehrenfest's theorem, 
and a conserved probability current density\cite{Nagao:2012mj,Nagao:2012ye}.

In the future-included CAT, 
let us consider the other 
ensemble\footnote{Just as each $| A_i (t) \rangle$ does not have to be orthogonal to each other, 
neither does each $| B_i (t) \rangle$.}  $\left\{ | B_i (t) \rangle \right\}$ 
besides the ensemble $\left\{ | A_i (t) \rangle \right\}$.  
Now we have a simple question: what kind of mixed states can be considered in 
the future-included theory? 
One possible candidate would be the same type of mixed states as we considered 
in the previous section. 
Since $| A_i (t) \rangle$ time-develops according to Eq.(\ref{Schr_for_Ai}) 
in the same way as before, let us consider 
the same mixed state described by the density matrix 
$\hat{\rho}_Q^{AA,\text{mixed}}(t)$ defined in Eq.(\ref{rhoQAmixed}) for $| A_i (t) \rangle$, 
and consider similar ones for $| B_i (t) \rangle$. 
Let us introduce a normalized state 
and an expectation value of an operator ${\cal O}$ for it by $| B_i (t) \rangle_{N} 
\equiv \frac{1}{\sqrt{ \langle B_i (t) |_Q ~B_i (t) \rangle} } | B_i (t) \rangle$ and  
$\langle \hat{\cal O} \rangle_Q^{B_i B_i} (t) 
\equiv  {}_{N} \langle B_i (t) |_Q {\cal O} | B_i (t) \rangle_{N}$, 
which time-develop as 
$i\hbar \frac{\partial}{ \partial t} | B_i (t) \rangle_{N} 
= \hat{H}_{Qh} | B_i (t) \rangle_{N} 
- \hat{\Delta} \left( \hat{H}_{Qa} ; | B_i (t) \rangle_{N} \right) | B_i (t) \rangle_{N}$, 
$\frac{d}{dt}\langle \hat{\cal O} \rangle_Q^{B_i B_i} (t)
= -\frac{i}{\hbar} \langle \left[ \hat{\cal O} , \hat{H}_{Qh} \right] \rangle_Q^{B_i B_i} (t) 
+\frac{i}{\hbar} \langle  \left\{  \hat{\cal O}, \hat{\Delta} \left( \hat{H}_{Qa} ; | B_i (t) \rangle_{N} \right) \right\} \rangle_Q^{B_i B_i} (t)$. 
Next let us consider 
a mixed state that is given by $| B_i (t) \rangle_{N}$ 
with the probability $r_i$ for each index $i$ ($r_i \ge 0$, $\sum_i r_i=1$).  
We define the density matrix to describe the mixed state 
and the expectation value of ${\cal O}$ for it by 
$\hat{\rho}_Q^{BB,\text{mixed}}(t) 
\equiv \sum_i r_i  | B_i (t) \rangle_N {}_N \langle B_i (t) |_Q 
\equiv \sum_i r_i \hat{\rho}_Q^{B_i B_i}(t)$, 
$\langle \hat{\cal O} \rangle_{\hat{\rho}_Q^{BB,\text{mixed}}} (t)
\equiv
\tr\left( \hat{\rho}_Q^{BB,\text{mixed}}(t)  \hat{\cal O}   \right) 
\equiv \sum_i r_i \langle \hat{\cal O} \rangle_{\hat{\rho}_Q^{B_i B_i}} (t) 
=\sum_i r_i \langle \hat{\cal O} \rangle_Q^{B_i B_i} (t)$, 
where $\hat{\rho}_Q^{B_i B_i}(t)$ obeys $\hat{\rho}_Q^{B_i B_i}(t)^2 = \hat{\rho}_Q^{B_i B_i}(t)$ 
and $\tr \left(\hat{\rho}_Q^{B_i B_i} (t) \right)=1$, so 
$\tr\left(\hat{\rho}_Q^{B B,\text{mixed}}(t) \right)=1$. 
We note that $\hat{\rho}_Q^{B B,\text{mixed}}(t)$ and 
$\langle \hat{\cal O} \rangle_{\hat{\rho}_Q^{BB,\text{mixed}}} (t)$ 
are $Q$-Hermitian and 
real for $Q$-Hermitian $\hat{\cal O}$, respectively. 
They time-develop as follows: 
$\frac{d}{dt} \hat{\rho}_Q^{BB,\text{mixed}} (t) 
=-\frac{i}{\hbar} \left[ \hat{H}_{Qh}, \hat{\rho}_Q^{BB,\text{mixed}} (t) \right] 
+\frac{i}{\hbar} \sum_i r_i $ 
$\left\{ \hat{\Delta} \left( \hat{H}_{Qa} ; 
| B_i (t) \rangle_{N} \right), 
\hat{\rho}_Q^{B_i B_i}  \right\} (t)$, 
$\frac{d}{dt} \langle \hat{\cal O} \rangle_{\hat{\rho}_Q^{BB,\text{mixed}}} (t) 
= 
-\frac{i}{\hbar} \langle \left[ \hat{\cal O} , \hat{H}_{Qh} \right] \rangle_{\hat{\rho}_Q^{BB,\text{mixed}}} (t) 
+\frac{i}{\hbar} \sum_i r_i$ \\
$\langle  \left\{  \hat{\cal O}, \hat{\Delta} \left( \hat{H}_{Qa} ; | B_i (t) \rangle_{N} \right)  \right\} \rangle_{\hat{\rho}_Q^{B_i B_i}} (t)$,  
which are almost the same as those for $\hat{\rho}_Q^{AA,\text{mixed}}(t)$. 
The only difference is that the sign in front of $\hat{H}_{Qa}$ is opposite. 
Therefore, if we use the automatic hermiticity mechanism 
for $| B_i (t) \rangle = \sum_j b_j^{(i)} (t) | \lambda_j \rangle$, then obtaining 
$| B_i(t) \rangle 
\simeq 
e^{ \frac{1}{\hbar} B (T_B-t)}  e^{-\frac{i}{\hbar} \hat{H}_{\text{eff}} (t-T_B)} 
| \tilde{B}_i (T_B) \rangle 
=\sum_{j \in A}  b_j^{(i)} (t) | \lambda_j \rangle 
\equiv | \tilde{B}_i (t) \rangle$ and 
$| B_i (t) \rangle_{N} 
\simeq \frac{1}{\sqrt{ \langle \tilde{B}_i (t) |_Q ~\tilde{B}_i(t) \rangle} } | \tilde{B}_i (t) \rangle 
\equiv | \tilde{B}_i (t) \rangle_{N}$ for large $T_B -t$, 
we find that the various relations for 
$\hat{\rho}_Q^{BB,\text{mixed}}(t) \simeq \hat{\rho}_Q^{\tilde{B}\tilde{B},\text{mixed}}(t)$ 
become the same as those for $\hat{\rho}_Q^{\tilde{A}\tilde{A},\text{mixed}}(t)$. 
Thus we have seen that the the automatic hermiticity mechanism works for mixed states 
described by the density matrices 
$\hat{\rho}_Q^{AA,\text{mixed}} (t) \simeq \hat{\rho}_Q^{\tilde{A} \tilde{A}, \text{mixed}} (t)$ and 
$\hat{\rho}_Q^{BB,\text{mixed}} (t) \simeq \hat{\rho}_Q^{\tilde{B} \tilde{B}, \text{mixed}} (t)$, 
and that, via the mechanism, both of the density matrices nicely obey the von Neumann equation 
with the effectively obtained $Q$-Hermitian Hamiltonian $\hat{H}_{\text{eff}}$. 
In addition, 
$\hat{\rho}_Q^{A_i A_i} (t)$ and $\hat{\rho}_Q^{B_i B_i} (t)$ 
have real meanings as density matrices of $| A_i (t) \rangle_N$ 
and $| B_i (t) \rangle_N$. 
However, neither 
$\tr\left( \hat{\rho}_Q^{A_i A_i} (t) \hat{\cal O} \right) 
={}_{N} \langle A_i (t) |_Q  \hat{\cal O} | A_i (t) \rangle_{N}$
nor 
$\tr\left( \hat{\rho}_Q^{B_i B_i} (t) \hat{\cal O} \right) 
={}_{N} \langle B_i (t) |_Q  \hat{\cal O} | B_i (t) \rangle_{N}$
matches the normalized matrix element $\langle \hat{\mathcal O} \rangle_Q^{B_i A_i} (t)$ 
given in Eq.(\ref{OQB_iA_i}). 
In the future-included CAT, we have a philosophy such that 
it is not ${}_{N} \langle A_i (t) |_Q  \hat{\cal O} | A_i (t) \rangle_{N}$ nor 
${}_{N} \langle B_i (t) |_Q  \hat{\cal O} | B_i (t) \rangle_{N}$ 
but $\langle \hat{\mathcal O} \rangle_Q^{B_i A_i} (t)$ that has a role of an expectation value of 
$\hat{\mathcal O}$. 
Therefore, $\hat{\rho}_Q^{A_i A_i} (t)$ and $\hat{\rho}_Q^{B_i B_i} (t)$ might 
not be good density matrices in this sense.  
Then what should we adopt as a density matrix 
in the future-included CAT if we wish to respect the philosophy?

We are now motivated to consider the other kind of density matrix 
such that 
the trace of the product of each component with an index $i$ and $\hat{\cal O}$ 
corresponds to $\langle \hat{\mathcal O} \rangle_Q^{B_i A_i} (t)$. 
Introducing 
$| A_i (t) \rangle_M \equiv 
\frac{| A_i (t) \rangle }{ \sqrt{ \langle B_i (t) |_Q A_i (t) \rangle} }$ and 
$| B_i (t) \rangle_M \equiv 
\frac{| B_i (t) \rangle }{ \sqrt{ \langle A_i (t) |_Q B_i (t) \rangle} }$,     
which obey 
$i \hbar \frac{d}{dt} | A_i (t) \rangle_M = \hat{H} | A _i (t) \rangle_M$, 
$i \hbar \frac{d}{dt} | B_i (t) \rangle_M = {\hat{H}}^{\dag^Q} | B _i (t) \rangle_M$,  
and 
${}_M\langle B_i (t) |_Q  A _i (t) \rangle_M = 1$, 
let us define the following ``skew density matrix'' $\hat{\rho}_Q^{BA,\text{mixed}}(t)$ 
and ``expectation value'' of $\hat{\cal O}$ 
for it\footnote{
$\langle \hat{\cal O} \rangle_{\hat{\rho}_Q^{BA,\text{mixed}}} (t) 
= \sum_i s_i \tr\left( \hat{\rho}_Q^{B_i A_i} (t) \hat{\cal O} \right) 
= \sum_i s_i 
\frac{\tr \left( \hat{\rho}_Q^{B_i B_i} (t)  \hat{\cal O}    \hat{\rho}_Q^{A_i A_i} (t) \right)}{
\tr \left( \hat{\rho}_Q^{B_i B_i} (t) \hat{\rho}_Q^{A_i A_i} (t) \right)}$ for $Q=1$ 
corresponds to the weak value for the generalized state introduced in Ref.~\cite{generalized_two-state_vector_formalism}, but is different from the generalized weak value 
$\frac{\tr(\hat{\rho}_f \hat{\cal O}  \hat{\rho}_i )}{\tr(\hat{\rho}_f  \hat{\rho}_i )}$ 
introduced in Refs.\cite{WuMolmer,TamateNakanishiKitano}.
The latter expression is more general since the numbers of ensembles of initial and final states 
for the density matrices $\hat{\rho}_i$ and $\hat{\rho}_f$ are taken independently, 
while, in our skew density matrix, the numbers of ensembles are supposed to be equal. 
This is because we are keeping in mind the maximization principle, by which a pair of initial and final 
states is generically chosen such that the absolute value of the transition amplitude is maximized.  
Then, in a situation such that a pair $\left\{ |A_i \rangle, | B_i \rangle \right\}$ 
and each weight $\left\{ s_i  \right\}$ are given, our skew density matrix enables us to 
calculate and simulate the ``expectation value" of ${\cal O}$.   
} by 
\begin{eqnarray}
\hat{\rho}_Q^{BA,\text{mixed}}(t) 
&\equiv&  \sum_i s_i  | A_i (t) \rangle_M {}_M\langle B_i (t) |_Q 
\equiv \sum_i s_i \hat{\rho}_Q^{B_i A_i} (t) , \\
\langle \hat{\cal O} \rangle_{\hat{\rho}_Q^{BA,\text{mixed}}} (t)
&\equiv&
\tr\left( \hat{\rho}_Q^{BA,\text{mixed}}(t)  \hat{\cal O}   \right) 
\equiv\sum_i s_i \langle \hat{\cal O} \rangle_{\hat{\rho}_Q^{B_i A_i}} (t) 
=\sum_i s_i \langle \hat{\cal O} \rangle_Q^{B_i A_i} (t) , 
\end{eqnarray}
where the weight $s_i$ for each $\hat{\rho}_Q^{B_i A_i} (t)$ obeys $s_i \ge 0$, $\sum_i s_i=1$, 
and $\hat{\rho}_Q^{B_i A_i} (t)$ obeys 
$\tr\left( \hat{\rho}_Q^{B_i A_i} (t)  \right) =1$, 
$\left( \hat{\rho}_Q^{B_i A_i} (t)  \right)^2 = \hat{\rho}_Q^{B_i A_i} (t)$. 
So $\tr\left(\hat{\rho}_Q^{BA,\text{mixed}} (t) \right)=1$. 
In addition, $\hat{\rho}_Q^{BA,\text{mixed}}(t)$ can be expressed as 
$\hat{\rho}_Q^{BA,\text{mixed}}(t)=\hat{U} (t-t_r) \hat{\rho}_Q^{BA,\text{mixed}} (t_r) \hat{U} (t-t_r) ^{-1}$, where  
$\hat{U} (t-t_r) \equiv e^{-\frac{i}{\hbar} \hat{H} (t-t_r)}$ is neither unitary nor $Q$-unitary, 
and $t_r$ is a reference time. 
They time-develop as follows: 
$\frac{d}{dt} \hat{\rho}_Q^{BA,\text{mixed}}(t) 
= 
-\frac{i}{\hbar} \left[ \hat{H}, \hat{\rho}_Q^{BA,\text{mixed}}(t)  \right]$, 
$\frac{d}{dt} \langle \hat{\cal O} \rangle_{\hat{\rho}_Q^{BA,\text{mixed}}} (t)
= 
-\frac{i}{\hbar} \langle \left[ \hat{\cal O} , \hat{H} \right] \rangle_{\hat{\rho}_Q^{BA,\text{mixed}}} (t)$, 
which show that $\hat{\rho}_Q^{BA,\text{mixed}} (t)$ obeys 
the von Neumann equation and Ehrenfest's theorem holds as they are. 
These properties are quite in contrast to those of  $\hat{\rho}_Q^{AA,\text{mixed}} (t)$ 
and $\hat{\rho}_Q^{BB,\text{mixed}} (t)$. 
If we consider the long time development, then for 
$| A_i (t) \rangle_M 
\simeq | \tilde{A}_i (t) \rangle_M 
\equiv \frac{| \tilde{A}_i (t) \rangle }{ \sqrt{ \langle \tilde{B}_i (t) |_Q \tilde{A}_i (t) \rangle} }$ and 
$| B_i (t) \rangle_M 
\simeq | \tilde{B}_i (t) \rangle_M 
\equiv \frac{| \tilde{B}_i (t) \rangle }{ \sqrt{ \langle \tilde{A}_i (t) |_Q \tilde{B}_i (t) \rangle} }$, 
we find that $\hat{\rho}_Q^{BA,\text{mixed}}(t)  \simeq \hat{\rho}_Q^{\tilde{B} \tilde{A},\text{mixed}}(t) 
\equiv \sum_i s_i  | \tilde{A}_i (t) \rangle_M {}_M\langle \tilde{B}_i (t) |_Q 
\equiv \sum_i s_i \hat{\rho}_Q^{\tilde{B}_i \tilde{A}_i}(t)$ 
and 
$\langle \hat{\cal O} \rangle_{\hat{\rho}_Q^{BA,\text{mixed}}} (t)
\simeq \langle \hat{\cal O} \rangle_{\hat{\rho}_Q^{\tilde{B} \tilde{A},\text{mixed}}} (t)
\equiv \tr\left(\hat{\rho}_Q^{\tilde{B} \tilde{A},\text{mixed}}(t)  \hat{\cal O}   \right)$  
$\equiv\sum_i s_i \langle \hat{\cal O} \rangle_{\hat{\rho}_Q^{\tilde{B}_i \tilde{A}_i}} (t) 
=\sum_i s_i \langle \hat{\cal O} \rangle_Q^{\tilde{B}_i \tilde{A}_i} (t)$ 
time-develop with an effectively obtained $Q$-Hermitian Hamiltonian $\hat{H}_{\text{eff}}$ 
as follows: 
$\frac{d}{dt} \hat{\rho}_Q^{\tilde{B} \tilde{A},\text{mixed}}(t)  
= -\frac{i}{\hbar} \left[ \hat{H}_{\text{eff}} , \hat{\rho}_Q^{\tilde{B} \tilde{A},\text{mixed}}(t) \right]$, 
$\frac{d}{dt} \langle \hat{\cal O} \rangle_{\hat{\rho}_Q^{\tilde{B} \tilde{A},\text{mixed}}} (t)
= -\frac{i}{\hbar}  \langle  \left[  \hat{\cal O}, \hat{H}_{\text{eff}} \right] \rangle_{\hat{\rho}_Q^{\tilde{B} \tilde{A},\text{mixed}}} (t)$. 
However, 
$\hat{\rho}_Q^{\tilde{B} \tilde{A},\text{mixed}}(t)$ and  
$\langle \hat{\cal O} \rangle_{\hat{\rho}_Q^{\tilde{B} \tilde{A},\text{mixed}}} (t)$ 
are neither $Q$-Hermitian nor real for $Q$-Hermitian $\hat{\cal O}$, respectively, 
because $| \tilde{A}_i (t) \rangle_M$ and $| \tilde{B}_i (t) \rangle_M$ are different states. 
This is quite in contrast to the cases for $\hat{\rho}_Q^{AA,\text{mixed}} (t)$ and 
$\hat{\rho}_Q^{BB,\text{mixed}} (t)$, 
where only either $| A_i (t) \rangle_N$ or  $| B_i (t) \rangle_N$ is used. 
To resolve this problem, we will consider it in another way.

\section{Hermiticity and reality for $\hat{\rho}_Q^{BA,\text{mixed}}(t)$ and $\langle \hat{\cal O} \rangle_{\hat{\rho}_Q^{BA,\text{mixed}}}(t)$}

In Ref.\cite{Nagao:2012mj}, utilizing the automatic hermiticity mechanism 
for pure states time-developing forward and backward,  
we obtained the following correspondence: 
\begin{equation}
\langle {\cal O} \rangle^{BA} ~\text{for large  $T_B-t$ and large $t- T_A$}
\simeq 
\langle {\cal O} \rangle_{Q'}^{AA} ~\text{for large $t- T_A$}, 
\end{equation}
based on the Schr\"{o}dinger equations (\ref{Schr_for_Ai}) 
and $i\hbar \frac{d}{dt} | B(t) \rangle = H^\dag| B(t) \rangle$, 
where $\langle {\cal O} \rangle^{BA}=\frac{\langle B(t) |  {\cal O} | A(t) \rangle}{\langle B(t) | A(t) \rangle}$ is a matrix element of an operator ${\cal O}$ defined with a usual inner product ($Q=1$) 
in the future-included theory, 
while $\langle {\cal O} \rangle_{Q'}^{AA}=\frac{\langle A(t) |_{Q'}  {\cal O} | A(t) \rangle}{\langle A(t) |_{Q'} A(t) \rangle}$ is a usual expectation value of ${\cal O}$ 
defined with a modified inner product $I_{Q'}$ in the future-not-included theory. 
We showed this correspondence by improving the method used in Ref.\cite{Bled2006}, 
which first multiplies $\langle {\cal O} \rangle^{BA}$ by 
$1=\frac{ \langle A(t) | B(t) \rangle }{\langle A(t) | B(t) \rangle}$ and then 
evaluates $| B(t) \rangle \langle B(t) |$. 
This correspondence strongly suggests that the future-included CAT is not excluded phenomenologically, 
even though it looks very exotic. 
Utilizing this method, let us estimate $\langle {\cal O} \rangle_Q^{BA}$ and 
$\hat{\rho}_Q^{BA}(t)$, 
based on the Schr\"{o}dinger equations (\ref{Schr_for_Ai}) 
and $i\hbar \frac{d}{dt} | B(t) \rangle = H^{\dag^Q} | B(t) \rangle$. 
Respecting the inner product $I_Q$ for all quantities, 
let us multiply $\langle {\cal O} \rangle_Q^{BA} (t)$ by 
$1=\frac{ \langle A(t) |_{Q} B(t) \rangle }{\langle A(t) |_{Q} B(t) \rangle}$, instead of 
$1=\frac{ \langle A(t) | B(t) \rangle }{\langle A(t) | B(t) \rangle}$. 
Then $\langle {\cal O} \rangle_Q^{BA}$ is rewritten as 
$\langle {\cal O} \rangle_Q^{BA} (t)
= \frac{ \langle A(t) |_{Q} B(t) \rangle   \langle B(t) |_Q  {\cal O} | A(t) \rangle}{\langle A(t) |_{Q} B(t) \rangle \langle B(t) |_Q A(t) \rangle}$. 
The expansion of $| B(T_B) \rangle$ used in Ref.\cite{Nagao:2012mj}, 
$| B(T_B) \rangle = \sum_i b_i | \lambda_i \rangle_B$ in terms of 
the eigenstate of $\hat{H}^\dag$, $|\lambda \rangle_B=Q|\lambda \rangle$, 
is found to produce too many $Q$, so it does not seem to be 
an appropriate choice in the present study. 
Hence we adopt another expansion: 
$| B(T_B) \rangle = \sum_i c_i | \lambda_i \rangle = \sum_i J(\lambda_i)^* | \lambda_i \rangle$, 
where $J(\lambda_i)$ is a function of $\lambda_i$. 
Then $| B(t) \rangle \langle B(t) |_Q$ is evaluated as follows: 
\begin{eqnarray}
| B(t) \rangle \langle B(t) |_Q
&=&  
e^{-\frac{i}{\hbar} \hat{H}^{\dag^Q} (t - T_B)} | B(T_B) \rangle  
\langle B(T_B) |_Q  e^{\frac{i}{\hbar} \hat{H} (t - T_B) }  \nonumber \\ 
&=&
\sum_{i,j} c_i  c_j^* e^{\frac{i}{\hbar}  \text{Re}(\lambda_j - \lambda_i) (t - T_B)} 
e^{ \frac{1}{\hbar} \text{Im}(\lambda_j + \lambda_i) (T_B - t)} 
| \lambda_i \rangle \langle \lambda_j |_Q  \nonumber \\ 
&\simeq&  
\frac{\int_{t-\Delta t}^{t+ \Delta t} | B(t) \rangle \langle B(t) |_Q dt }{ \int_{t-\Delta t}^{t+ \Delta t} dt } 
\simeq
\sum_i | c_i |^2  e^{ \frac{2}{\hbar}\text{Im} \lambda_i (T_B - t)}  | \lambda_i \rangle ~\langle \lambda_i |_Q  
\nonumber \\
&\simeq& e^{ \frac{2}{\hbar}B (T_B - t)}  Q_4  \quad  \text{for large $T_B - t$} , 
\end{eqnarray}
where in the third line we have smeared the present time $t$ a little bit, 
and the off-diagonal elements wash to $0$. 
In the last line we have used the automatic hermiticity mechanism  
for large $T_B - t$, and introduced  
$Q_4 \equiv \sum_{i \in A} | c_i |^2  | \lambda_i \rangle \langle \lambda_i |_Q$, 
which is expressed as follows: 
$Q_4
=J(\hat{H}_{\text{eff}} + i B \Lambda_A )^{\dag^Q} 
\Lambda_A 
J(\hat{H}_{\text{eff}} + iB \Lambda_A ) 
=Q^{-1} \tilde{J}(\hat{H}_{\text{eff}} )^\dag  Q \tilde{J}(\hat{H}_{\text{eff}} ) 
\equiv Q^{-1} Q_{\tilde{J}}$. 
Here, supposing that $\text{Re} \lambda_i$ are not degenerate, 
we have introduced 
$\Lambda_A \equiv \sum_{i \in A} | \lambda_i \rangle \langle \lambda_i |_Q$, 
a function $\tilde{J}$ 
such that $\tilde{J}(\text{Re} \lambda_i ) \equiv J(\text{Re} \lambda_i + iB)  = c_i^*$ for $i \in A$, 
and $Q_{\tilde{J}} \equiv \tilde{J}(\hat{H}_{\text{eff}} )^\dag  Q \tilde{J}(\hat{H}_{\text{eff}} )$. 
Now we use the automatic hermiticity mechanism for large $t-T_A$.   
Then, since $| A(t) \rangle \equiv \sum_i  a_i(t) | \lambda_i \rangle $ behaves as 
$| \tilde A(t) \rangle \equiv \sum_{i \in A}  a_i(t) | \lambda_i \rangle$, 
we obtain 
$\langle {\cal O} \rangle_Q^{BA} 
\simeq
\frac{  \langle \tilde{A} (t) |_{Q_{\tilde{J}}}  {\cal O}  | \tilde{A}(t) \rangle }
{ \langle \tilde{A}(t) |_{Q_{\tilde{J}}}  \tilde{A}(t) \rangle } 
\equiv \langle {\cal O} \rangle_{Q_{\tilde{J}}}^{\tilde{A}\tilde{A}} 
\quad \text{for large $T_B - t$ and large $t-T_A$}$. 
Next, let us consider the expectation value in the future-not-included theory: 
$\langle {\cal O} \rangle_{Q_J}^{AA} 
\equiv
\frac{  \langle A(t) |_{Q_J} {\cal O}  | A(t) \rangle }{ \langle A(t) |_{Q_J} A(t) \rangle }$,  
where $Q_J \equiv J(\hat{H})^\dag Q J(\hat{H}) =({P_{J^{-1}} }^{-1} )^\dag {P_{J^{-1}} }^{-1}$, 
and $P_{J^{-1}} \equiv J(\hat{H})^{-1} P$ diagonalizes $\hat{H}$: 
$(P_{J^{-1}})^{-1} \hat{H} P_{J^{-1}} = P^{-1} \hat{H} P = D$.  
We introduce 
$| \lambda_i \rangle^{J^{-1}} \equiv J(\hat{H})^{-1} | \lambda_i \rangle$, 
so that $| \lambda_i \rangle^{J^{-1}}$ is $Q_J$-orthogonal, i.e., 
$I_{Q_J} (   | \lambda_i \rangle^{J^{-1}} ,  | \lambda_j \rangle^{J^{-1}} ) 
\equiv {}^{J^{-1}}\langle \lambda_i | Q_J | \lambda_j \rangle^{J^{-1}} 
=\delta_{ij}$. 
We use the automatic hermiticity mechanism for large $t-T_A$. 
$| A(t) \rangle$ behaves as 
$| \tilde A(t) \rangle = \sum_{i \in A}  a_i(t) | \lambda_i \rangle$, 
and $Q_J$ is estimated as follows: 
$Q_J 
\simeq  J(\hat{H}_{\text{eff}} + iB \Lambda_A )^\dag Q J(\hat{H}_{\text{eff}} + iB \Lambda_A ) 
=\tilde{J}(\hat{H}_{\text{eff}} )^\dag Q \tilde{J}(\hat{H}_{\text{eff}} ) 
= Q_{\tilde{J}}$.   
Then the expectation value in the future-not-included theory is expressed as 
$\langle {\cal O} \rangle_{Q_J}^{AA} 
\simeq \frac{  \langle \tilde{A}(t) |_{Q_{\tilde{J}}} {\cal O}  | \tilde{A}(t) \rangle }
{ \langle \tilde{A}(t) |_{Q_{\tilde{J}}} \tilde{A}(t) \rangle } 
=\langle {\cal O} \rangle_{Q_{\tilde{J}}}^{\tilde{A}\tilde{A}} \quad \text{for large $t-T_A$}$.  
Thus we have obtained the following correspondence: 
\begin{equation}
\langle {\cal O} \rangle_Q^{B A} ~\text{for large $T_B-t$ and large $t-T_A$}
\quad 
\simeq 
\quad
\langle {\cal O} \rangle_{Q_{\tilde{J}}}^{\tilde{A}\tilde{A}}
\quad
\simeq
\quad
\langle {\cal O} \rangle_{Q_J}^{AA} ~\text{for large $t-T_A$}, 
\end{equation}
which suggests that 
the future-included theory is not excluded, although it looks very exotic. 
$\langle {\cal O} \rangle_{Q_{\tilde{J}}}^{\tilde{A}\tilde{A}}$ is real 
for $Q_{\tilde{J}}$-Hermitian ${\cal O}$, and time-develops 
according to the $Q_{\tilde{J}}$-Hermitian Hamiltonian $\hat{H}_{\text{eff}}$. 
We can apply this correspondence to each $i$-component 
$\langle \hat{\cal O} \rangle_Q^{B_i A_i} (t)$.

Next let us evaluate the skew density matrix 
$\hat{\rho}_Q^{B A} (t)=\frac{| A (t) \rangle \langle B (t) |_Q }{\langle B (t) |_Q A (t) \rangle }$ 
by multiplying it by $1=\frac{ \langle A(t) |_{Q} B(t) \rangle }{\langle A(t) |_{Q} B(t) \rangle}$. 
Utilizing the above evaluation of $| B(t) \rangle \langle B(t) |_Q$, 
we obtain the correspondence: 
\begin{equation}
\hat{\rho}_Q^{B A} (t) ~\text{for large $T_B-t$ and large $t-T_A$}
\quad 
\simeq 
\quad
\hat{\rho}_{Q_{\tilde{J}}}^{\tilde{A}\tilde{A}} (t)
\quad
\simeq   \quad 
\hat{\rho}_{Q_J}^{AA}  (t)~\text{for large $t-T_A$} , 
\end{equation}
where 
$\hat{\rho}_{Q_{\tilde{J}}}^{\tilde{A}\tilde{A}}(t)
\equiv 
\frac{| \tilde{A} (t) \rangle \langle \tilde{A}(t) |_{Q_{\tilde{J}}}   }{\langle \tilde{A}(t) |_{Q_{\tilde{J}}} \tilde{A} (t) \rangle }  
=\hat{U}_{\text{eff}} (t-t_r) 
\hat{\rho}_{Q_{\tilde{J}}}^{\tilde{A}\tilde{A}}(t_r) 
\hat{U}_{\text{eff}} (t-t_r)^{\dag^{Q_{\tilde{J}}}}$. 
Here $t_r$ is a reference time, and $\hat{\rho}_{Q_{\tilde{J}}}^{\tilde{A}\tilde{A}}(t)$ obeys 
$\tr\left( \hat{\rho}_{Q_{\tilde{J}}}^{\tilde{A}\tilde{A}} \right) =1$ and 
$\left( \hat{\rho}_{Q_{\tilde{J}}}^{\tilde{A}\tilde{A}} \right)^2 = \hat{\rho}_{Q_{\tilde{J}}}^{\tilde{A}\tilde{A}}$.
$\hat{U}_{\text{eff}} (t-t_r) = e^{-\frac{i}{\hbar} \hat{H}_{\text{eff}} (t-t_r)}$ is $Q_{\tilde{J}}$-unitary, 
and $\hat{\rho}_{Q_{\tilde{J}}}^{\tilde{A}\tilde{A}}(t)$ is $Q_{\tilde{J}}$-Hermitian. 
We can apply this correspondence to each $i$-component 
$\hat{\rho}_Q^{B_i A_i} (t)$. 
Therefore, though our skew density matrix $\hat{\rho}_Q^{B_i A_i} (t)$ 
is not $Q$-Hermitian by its definition, 
after a long time development it results in a usual expression of density matrix 
$\hat{\rho}_{Q_{\tilde{J}}}^{\tilde{A}_i \tilde{A}_i} (t)$ 
that is $Q_{\tilde{J}}$-Hermitian. 
Application to $\hat{\rho}_Q^{BA,\text{mixed}}(t) = \sum_i s_i \hat{\rho}_Q^{B_i A_i} (t)$ 
is rather straightforward and we easily see that it time-develops similarly. 
Indeed, applying this correspondence to each component 
$\hat{\rho}_Q^{B_i A_i} (t)$, 
we find that the expectation value of ${\cal O}$ for $\hat{\rho}_Q^{B_i A_i} (t)$, 
$\langle \hat{\cal O} \rangle_{\hat{\rho}_Q^{B_i A_i}} (t)$, is expressed 
for large $T_B-t$ and large $t-T_A$ as  
\begin{eqnarray}
&&\langle \hat{\cal O} \rangle_{\hat{\rho}_Q^{B_i A_i}} (t) 
=
\tr\left( \hat{\rho}_Q^{B_i A_i} (t) \hat{\cal O} \right) 
\simeq 
\tr\left( \hat{\rho}_{Q_{\tilde{J}}}^{\tilde{A}_i \tilde{A}_i} (t) \hat{\cal O} \right) 
\equiv
\langle \hat{\cal O} \rangle_{\hat{\rho}_{Q_{\tilde{J}}}^{\tilde{A}_i \tilde{A}_i}} (t) 
=\langle {\cal O} \rangle_{Q_{\tilde{J}}}^{\tilde{A}_i \tilde{A}_i} (t) , 
\end{eqnarray}
which is real for $Q_{\tilde{J}}$-Hermitian ${\cal O}$. 
Finally,  
$\hat{\rho}_Q^{BA,\text{mixed}}(t)  \simeq \hat{\rho}_{Q_{\tilde{J}}}^{\tilde{A} \tilde{A},\text{mixed}}(t) 
= \sum_i s_i \hat{\rho}_{Q_{\tilde{J}}}^{\tilde{A}_i \tilde{A}_i}(t)$ 
and 
$\langle \hat{\cal O} \rangle_{\hat{\rho}_Q^{BA,\text{mixed}}} (t)
\simeq \langle \hat{\cal O} \rangle_{\hat{\rho}_{Q_{\tilde{J}}}^{\tilde{A} \tilde{A},\text{mixed}}} (t)
=\sum_i s_i \langle \hat{\cal O} \rangle_{\hat{\rho}_{Q_{\tilde{J}}}^{\tilde{A}_i \tilde{A}_i}} (t)$
time-develop according to 
\begin{eqnarray} 
&&
\frac{d}{dt} \hat{\rho}_{Q_{\tilde{J}}}^{\tilde{A} \tilde{A},\text{mixed}}(t)
= 
-\frac{i}{\hbar} \left[ \hat{H}_{\text{eff}} , \hat{\rho}_{Q_{\tilde{J}}}^{\tilde{A} \tilde{A},\text{mixed}}(t) \right] , \\
&&
\frac{d}{dt} \langle \hat{\cal O} \rangle_{\hat{\rho}_{Q_{\tilde{J}}}^{\tilde{A} \tilde{A},\text{mixed}}} (t)
=
-\frac{i}{\hbar}  \langle  \left[  \hat{\cal O}, \hat{H}_{\text{eff}} \right] 
\rangle_{Q_{\tilde{J}}}^{\tilde{A} \tilde{A},\text{mixed}} (t) , 
\end{eqnarray}
which show that 
$\hat{\rho}_Q^{BA,\text{mixed}}(t)  \simeq \hat{\rho}_{Q_{\tilde{J}}}^{\tilde{A} \tilde{A},\text{mixed}}(t)$ 
obeys the von Neumann equation with 
the $Q_{\tilde{J}}$-Hermitian Hamiltonian $\hat{H}_{\text{eff}}$ and Ehrenfest's theorem holds. 
We note that $\hat{\rho}_{Q_{\tilde{J}}}^{\tilde{A} \tilde{A},\text{mixed}}(t)$ is 
$Q_{\tilde{J}}$-Hermitian, and 
$\langle \hat{\cal O} \rangle_{\hat{\rho}_{Q_{\tilde{J}}}^{\tilde{A} \tilde{A},\text{mixed}}} (t)$ 
is real for $Q_{\tilde{J}}$-Hermitian ${\cal O}$. 
Thus we have seen that the problem with $\hat{\rho}_Q^{BA,\text{mixed}}(t)$ and 
$\langle \hat{\cal O} \rangle_{\hat{\rho}_Q^{BA,\text{mixed}}}(t)$ 
mentioned at the end of the previous section can be effectively resolved 
by considering the long time development for large $T_B-t$ and large $t-T_A$.

\section{Discussion}

We first reviewed the modified inner product $I_Q$ that 
makes a given non-normal Hamiltonian normal 
with regard to it, 
and the automatic hermiticity mechanism\cite{Nagao:2010xu,Nagao:2011za,Nagao:2012mj}, which 
we previously proposed and studied 
for pure states in the CAT. 
Next, in the case of the future-not-included CAT, we defined 
a density matrix $\hat{\rho}_Q^{AA,\text{mixed}}(t)$ to describe a mixed state and 
an expectation value of an operator ${\cal O}$ for it, 
and studied their properties.  
In the classical limit, 
$\langle \hat{\cal O} \rangle_Q^{A_i A_i}(t)$ 
time-develops by a $Q$-Hermitian Hamiltonian, and Ehrenfest's theorem holds. 
In addition, we showed that, if we consider a long time development, 
eigenvectors having the largest imaginary part of the eigenvalues of $\hat{H}$ dominate most. 
On the subspace spanned by such eigenvectors, 
a $Q$-Hermitian Hamiltonian effectively emerges, the expectation value of 
${\cal O}$ becomes real for $Q$-Hermitian ${\cal O}$,   
and the density matrix obeys the von Neumann equation with the $Q$-Hermitian Hamiltonian.  
Thus we confirmed that the automatic hermiticity mechanism works for mixed states 
in the future-not-included theory.

The situation becomes more non-trivial in the future-included theory, 
because, 
in the future-included theory, there are two 
classes of 
ensembles of state vectors,  $\left\{ | A_i (t) \rangle \right\}$ and  $\left\{ | B_i (t) \rangle \right\}$,  
that time-develop 
forward from the initial time $T_A$ and backward from the final time $T_B$, respectively.  
So it seems that there are at least a couple of candidates for density matrices in the future-included theory. 
As the first candidate, we investigated a pair of density matrices 
$\hat{\rho}_Q^{AA,\text{mixed}}(t)$ and $\hat{\rho}_Q^{BB,\text{mixed}}(t)$, 
which are composed of only either 
$\left\{ | A_i (t) \rangle \right\}$ or $\left\{ | B_i (t) \rangle \right\}$, and argued that, 
though the pair has nice properties, 
it has a common disadvantage in the future-included theory. 
In general, the trace of the product of 
a density matrix and an operator ${\cal O}$ 
has to match an expectation value of ${\cal O}$, but this is not the case for this pair, 
because it is the matrix element of ${\cal O}$, $\langle {\cal O} \rangle_Q^{BA}$, 
that is expected to work as an expectation value of ${\cal O}$ in the future-included theory. 
To resolve this problem, we introduced a ``skew density matrix'' $\hat{\rho}_Q^{BA,\text{mixed}}(t)$, 
which is composed of both 
$\left\{ | A_i (t) \rangle \right\}$ and $\left\{ | B_i (t) \rangle \right\}$. 
The skew density matrix has a nice property such that the trace of the product of it 
and an operator ${\cal O}$ 
matches the matrix element $\langle {\cal O} \rangle_Q^{BA}(t)$.   
It also obeys the the von Neumann equation as it is. 
In addition, 
utilizing the automatic hermiticity mechanism, we showed that 
the skew density matrix $\hat{\rho}_Q^{BA,\text{mixed}}(t)$ 
and matrix element 
$\langle {\cal O} \rangle_Q^{BA}(t)$ defined with an inner product $I_Q$  
in the future-included theory for large $T_B-t$ and large $t-T_A$ approximately correspond to 
another density matrix $\hat{\rho}_{Q_J}^{AA,\text{mixed}}(t)$ and an expectation value 
$\langle {\cal O} \rangle_{Q_J}^{AA}(t)$ defined with another inner product $I_{Q_J}$ 
in the future-not-included theory for large $t-T_A$. 
Therefore, even though the skew density matrix is not a density matrix in a usual sense, 
it can effectively work as if it were a usual density matrix. 
Thus we argued that it is the skew density matrix that is expected to have a role of a density matrix 
in the future-included theory.  
In addition, we confirmed that the automatic hermiticity mechanism works for mixed states  
in the future-included theory.

Now density matrices have been implemented in the CAT. 
What should we study by using them?
It would be interesting to investigate the classical dynamics of the CAT in phase space 
via the Wigner function. 
For this purpose, it would be better to study further in detail 
the harmonic oscillator model that we previously formulated by introducing 
the two-basis formalism\cite{Nagao:2019dew}. 
Also, it would be intriguing to evaluate von Neumann entropy in the CAT. 
Furthermore, density matrices are necessary tools 
if we wish to investigate quantum measurement quantitatively 
in a composite system via the master equation.
We would like to report such studies in the future.



\ack

This work was supported by JSPS KAKENHI Grant Number JP21K03381, and accomplished 
during K.N.'s sabbatical stay in Copenhagen. 
He would like to thank the members and visitors of NBI 
for their kind hospitality and Klara Pavicic for her various generous 
considerations during his visits to Copenhagen.  
H.B.N. is grateful to NBI for allowing him to work there as emeritus. 
In addition, the authors would like to express their gratitude to David Gross 
for stimulating discussion at the Bohr-100 workshop, 
and to Klaus M$\o$lmer for intensive discussion on retrodictions.
Furthermore, they are grateful to the referee of the journal for giving them valuable comments.

\let\doi\relax


\end{document}